\documentclass[prl,aps,twocolumn, floatfix,preprintnumbers]{revtex4}

\usepackage[applemac]{inputenc}
\usepackage[T1]{fontenc} 
 \usepackage{amsmath} 
\usepackage{subfigure}
\usepackage{lipsum}
\usepackage{amsfonts} 
\usepackage{amssymb, mathrsfs}
\usepackage{braket}
\usepackage{graphicx} 
\usepackage[usenames]{color} 

\usepackage{bbm}

\def\beq{\begin{equation}}
\def\eeq{\end{equation}}
\def\bsp{\begin{split}}
\def\esp{\end{split}}
\def\bea{\begin{eqnarray}}
\def\eea{\end{eqnarray}}
\def\ba{\begin{array}}
\def\ea{\end{array}}

\def\lb{\left(}
\def\rb{\right)}

\def\l.{\left.}
\def\r.{\right.}

\def\part{\partial}

\def\bra#1{{\cal{h}} #1 \mid}
\def\ket#1{\mid #1 {\cal{i}}}

\makeatletter

\newcommand{\Rmnum}[1]{\expandafter\@slowromancap\romannumeral #1@}
\makeatother
\begin{document}
\title{Haldane-like antiferromagnetic spin chain in the large anisotropy limit}
\author{S. A. Owerre}
\author{M. B. Paranjape} 
\affiliation{Groupe de physique des particules, D\'epartement de physique,
Universit\'e de Montr\'eal,
C.P. 6128, succ. centre-ville, Montr\'eal, 
Qu\'ebec, Canada, H3C 3J7 }

\begin{abstract}

\section*{ABSTRACT}  
We consider the one dimensional, periodic spin chain with $N$ sites, similar to the one studied by Haldane \cite{hal}, however in the opposite limit of very large anisotropy and small nearest neighbour, anti-ferromagnetic exchange coupling between the spins, which are of large magnitude $s$.  For a  chain with an even number of sites we show that actually the ground state is non degenerate and given by a superposition of the two Néel states, due to quantum spin tunnelling.  With an odd number of sites, the Néel state must necessarily contain a soliton. The position of the soliton is arbitrary thus the  ground state is $N$-fold degenerate.  This set of  states reorganizes into a band. We show that this occurs at order $2s$ in perturbation theory.  The ground state is non-degenerate for integer spin, but degenerate for half-odd integer spin as is required by Kramer's theorem \cite{kram}.   
\end{abstract}



\maketitle


{\it{Introduction}-} The study of spin chains has attracted considerable attention in condensed matter and particle physics over the years.  The breakthrough in this subject was begun by the work of Bethe and Hulth\'en \cite{bethe} for one-dimensional $(D=1)$, isotropic Heisenberg spin-$\frac{1}{2}$ antiferromagnetic chain. They computed the exact antiferromagnetic ground state and its energy for an infinite chain.  Anderson \cite{B} had worked out the ground state energies and the spectrum for $D=1,2,3$ by means of spin wave theory. The inclusion of an anisotropy term introduces much interesting physics ranging from quantum computing \cite{loss} to optical physics \cite{simon}.   The resulting Hamiltonian now possesses two coupling constants which can compete against each other:
\bea
\hat H= -K\sum_{i=1}^{N} S_{i,z}^2 +\lambda\sum_{i=1}^{N} \vec S_i\cdot \vec S_{i+1}=\hat H_{0} +\hat V
\label{eqn4}
\eea
Each spin has magnitude $|\vec S_i|=s$ and we will consider the large $s$ limit.  The two limiting cases are weak  anisotropy $\lambda\gg K$ and weak exchange coupling  $\lambda\ll K$, where  $\lambda$ is the Heisenberg exchange interaction coupling constant  and $K$ is the anisotropy coupling constant.  The limit of weak anisotropy  was studied by Haldane \cite{hal} in a closely related model.  He demonstrated that in  the large spin limit, $s\gg1$, the system can be mapped to a non-linear sigma model in field theory with distinguishing effects between integer and half-odd integer spins.  In this letter we will also study the large spin limit, but take the opposite limit of strong anisotropy.

With $\lambda=0$, the ground state is $2^N$ fold degenerate, corresponding to each spin in the state $S_z=\pm s$.  For an even number of sites, the model is bi-partite, and the two fully anti-aligned Ne\'el states are good starting points to investigate the ground state.  For an odd number of sites, the Néel states are frustrated, they must  contain at least one defect, which are sometimes called domain wall solitons \cite{vill}. There is a high level of degeneracy as the soliton can be placed anywhere along the cyclic chain and this degenerate system is the starting point to investigate the ground state for the case of an odd number of sites.  Frustrated systems are of great importance in condensed matter physics as they lead to exotic phases of matter such as spin liquid\cite{bal}, spin classes\cite{bin} and topological orders \cite{kit}. Solitons will also occur on the periodic chain with even number of sites, but they must occur in soliton anti-soliton pairs.  Villain \cite{vill} has studied the one-dimensional $XXZ$ antiferromagnetic spin chain, however for spin-$\frac{1}{2}$  close to the Ising limit, where our analyses are quite parallel.
In this letter, we will study the  spin chain with Hamiltonian given by the simple form given in Eq.\eqref{eqn4}  with periodic boundary condition $\vec S_{N+1}=\vec S_1$, and we consider $K\gg \lambda>0$, {\sl i.e.} strong easy-axis anisotropy  and perturbative Heisenberg antiferromagnetic coupling. The interaction term is denoted by $\hat V$ and the free term is denoted by $\hat H_{0}$ in Eq.\eqref{eqn4}.

{\it{Spin wave theory}-} In our model, we expect the spin waves to have a large gap.  This is because a spin wave corresponds to introducing a local deviation of the spins away from their respective highest or lowest values of $S_z$.  This incurs energy cost controlled by the free Hamitonian $\hat H_0$.  Thus the energy cost is proportional to $K$.    Noticing that the classical ground states of our model are locally the fully anti-aligned Ne\'el states, we can introduce small quantum fluctuations in the spirit of Holstein-Primakoff transformation \cite{hh}. 
With a straightforward analysis we obtain that the magnon (spin wave) dispersion is given by \cite{hal}
\bea
\varepsilon _q = \bigg[\varepsilon _0^2+\varepsilon _1^2\sin^2(q)\bigg]^{1/2}
\label{mag}
\eea
where $\varepsilon _0=2s\sqrt{K(K+2\lambda)}$, $\varepsilon _1=2s\lambda$ and ${-\pi}\leq q \leq{\pi}$. As $K\neq 0$ and in fact large, the magnon dispersion has a large gap.


{\it{Even number of sites and spin coherent state path integral}-} 
First-order degenerate perturbation theory in  the interaction term requires that we  diagonalize the interaction Hamiltonian term in the $2^N$ degenerate subspace.  The interaction can be written as
\beq
\hat V=\lambda\sum_{i=1}^{N} \left[ S_i^z\cdot  S_{i+1}^z+\frac{1}{2}\left(S^+_iS^-_{i+1}+S^-_iS^+_{i+1}\right)\right]
\eeq
which has a diagonal term and a term which induces transitions of spin pairs to states which are no longer in the highest or lowest weight states.  Thus, at first-order, only the diagonal term in the interaction has a nonzero matrix element  within the degenerate subspace. The two Ne\'el states  have the lowest energy with $E_{AF}=(-K-{\lambda})Ns^2$  and the two fully aligned ferromagnetic states have the highest energy with $E_{F}=(-K+{\lambda})Ns^2$. The energies of the intermediate states lie between $E_{AF}$ and $E_{F}$. The energies of the two Ne\'el states are corrected in each order of perturbation theory, however the correction is identical for each, thus they remain degenerate.  However at order  $2s\lb\frac{N}{2}\rb$ the non-diagonal part of the interaction flips $N/2$ pairs of spins, causing each spin to reverse.  This transforms the Néel states into each other, and thus at this order in perturbation theory, we must diagonalize the effective Hamiltonian.  As there are only two degenerate states, call them $\ket{\pm p}$, this correspond simply to a $2\times 2$ matrix and the ground state energy splitting has the form \cite{bab1}
\bea
\Delta = 2\braket{-p|\hat V\mathcal{A}^{{sN-1}}|p}
\label{eq5}
\eea
where $\mathcal{A}^{{sN-1}}= \lb\frac{\mathcal{Q}}{E_{s}-\hat H_{0}}\hat V\rb^{sN-1} + O(\mathcal{Q}V)^{sN-2}$, $E_s=-KNs^2$ and $\mathcal{Q}=1-\ket{p}\bra{p}-\ket{-p}\bra{-p}$ is the operator projecting to the complement of the degenerate sub-space. With a bit of work, the energy splitting can be obtained from Eq.\eqref{eq5}, however we will obtain the desired  result via the spin coherent state path integral formalism \cite{AB,kla,ams}.
 In this formalism, each spin is represented by a unit vector, and the corresponding (Euclidean) Lagrangian is given by
\begin{align}
L_E&= is\sum_i \dot{\phi}_i(1-\cos\theta_i) +K\sum_i\sin^2\theta_i \nonumber\\&+\lambda\sum_i[\sin\theta_i\sin\theta_{i+1}\cos(\phi_i-\phi_{i+1})+\cos\theta_i\cos\theta_{i+1}].
\end{align}
The first term is the usual Wess-Zumino \cite{wznw} term which arises from the non-orthogonality of spin coherent states while the other two terms correspond to the anisotropy energy and the exchange energy.  Quantum amplitudes are obtained via the path integral.  Solutions of the (Euclidean) classical equations of motion give information about quantum tunnelling amplitudes.  The classical equation of motion for $\phi_i$ yields
\begin{align}
is\frac{d(1-\cos\theta_i)}{d\tau}&= \sin\theta_{i-1}\sin\theta_i\sin(\phi_{i-1}-\phi_i)\nonumber\\&-\sin\theta_{i}\sin\theta_{i+1}\sin(\phi_{i}-\phi_{i+1})
\end{align}
Similar expression holds for the equation of motion for $\theta_i$.
Summing both sides of this equation one obtains
\begin{align}
&is\sum_i\frac{d(1-\cos\theta_i)}{d\tau}=0\Rightarrow\sum_i\cos\theta_i=l=0
\label{eqn8}
\end{align}
which corresponds to the conservation of  $z$-component of the total spin $\sum_i S_{i}^z$, as the full Hamiltonian,  Eq.\eqref{eqn4},  is invariant under rotations about the $z$ axis. A particular solution of Eq.\eqref{eqn8} is $\theta_{2k-1}\equiv\theta$, and $\theta_{2k}=\pi-\theta$, $k=1,2\cdots, N$.  Hence the effective Lagrangian (adding an irrelevant constant) becomes
\begin{align}
L_E^{eff} &= is\sum_{k=1}^{N}{\dot \phi}_k-is\cos\theta\sum_{k=1}^{N/2}({\dot \phi}_{2k-1}-{\dot \phi}_{2k})\nonumber\\&+ \sum_{i=1}^{N}\bigg[K+\lambda[1+\cos(\phi_{i}-\phi_{i+1})]\bigg]\sin^2\theta\\&= isN\dot\Phi- \frac{isN}{2}\dot\phi\cos\theta+ U_{eff}\end{align}
where $ U_{eff}= N[K+\lambda(1+\cos\phi)]\sin^2\theta$ and the last equality is obtained by making the further simplifying ansatz $\phi_{i}-\phi_{i+1}=(-1)^{i+1}\phi$  effectively  reducing  to a single spin problem.   The instanton that we will find must go from $\theta=0$ to $\theta=\pi$.  Conservation of energy implies $\partial_\tau U_{eff}=0$, which then must vanish, $U_{eff}=0$, since it is so at $\theta=0$.  This implies
\bea
{\cos\phi} = -\lb\frac{K}{\lambda}+1\rb \ll 1
\eea
since $\sin\theta(\tau )\ne 0$ along the whole trajectory.
Thus $\phi$ is a complex constant which can be written as $\phi=\pi+i\phi_I$ similar to that of two spin case \cite{ams}. 
The classical equation of motion for $\phi$ gives
\bea
is\dot \theta= -2\lambda\sin\theta\sin\phi= i2{\lambda}\sin\theta\sinh\phi_I
\eea
which integrates as
\beq 
\theta\lb \tau\rb =  2 \arctan \lb e^{\omega (\tau-\tau_0)}\rb
\label{18}
\eeq
 where $\omega= (2\lambda/s)\sinh\phi_I$. The instanton is independent of the number of spins and only depends on the initial and the final points.  As found in \cite{ams} the instanton contributes to the action only through the Wess-Zumino term, as $U_{eff}=0$ all along the trajectory.  The action is given by\cite{ams}
 \bea
S_c&= &S_0-\frac{isN}{2}\int_0^{\pi+i\phi_I} \hskip-.8cm  d\phi \cos\theta|_{\theta=0} -\frac{isN}{2} \int^0_{\pi+i\phi_I}  \hskip-.8cm d\phi \cos\theta|_{\theta=\pi}\nonumber \\
&=&0-isN\pi +Ns\phi_I=-isN\pi +Ns\phi_I
\eea
The two Néel states reorganize into the symmetric and antsymmetric linear superpositions, $\ket+$ and the $\ket-$ as in \cite{ams}.  The energy splitting is then
\bea
 \Delta= 2\mathscr{D} e^{-S_c}=2\mathscr{D}\lb\frac{\lambda}{2K}\rb^{Ns}\cos(sN\pi)
 \label{eq15}
\eea
where $\mathscr{D}$ is a determinantal pre-factor which contains no $\lambda$ dependence. The factor of $\lambda^{Ns}$ signifies the order of degenerate perturbation theory as can be easily verified from Eq.\eqref{eq5}. The energy splitting, Eq.\eqref{eq15} is the general formula for any even spin $N$.  For $N=2$ we recover the results obtained previously\cite{bab1,ams}.  The factor $sN$ can be even or odd, depending on the value of the spin.  For half odd integer spin, and for $N=2(2k+1)$ we find $\Delta$ is negative which means that $\ket -$ is the ground state and $\ket +$ is the first excited state.  In all other cases, for any value of the spin $s$ and $N=2 (2k)$ we find $\Delta$ is positive and then $\ket +$ is the ground state, $\ket -$ is the first excited state.  

{\it  Odd spin chain, frustration and solitons-} When we consider a periodic chain with an odd  number of sites  a soliton like defect arises due to the spin frustration. The fully anti-aligned Néel like state cannot complete periodically, as it requires an even total number of spins.  Thus there has to be at least one pair of spins that is aligned.  This can come in the form up-up or down-down while all other pairs of neighbouring spins are in the up-down or down-up combination.  As the total $z$ component of the spin is conserved, these states lie in orthogonal super-selection sectors and never transform into each other.  The position of the soliton is arbitrary thus each sector is $N$-fold degenerate.  In the first case the total $z$ component of the spin is $s$ while in the second case it is $-s$.  We will without loss of generality consider the $s$ sector.   These degenerate states are denoted by $\ket k$, $k=1,\cdots,N$ where
\beq
\ket{k}=\ket{\uparrow, \downarrow,\uparrow,\downarrow,\uparrow, \cdots,\underbrace{\uparrow,\uparrow,}_{k,k+1^{th}\,{\rm place}},\cdots ,\uparrow, \downarrow}
\eeq
in obvious notation.  These states have the same energy $E_s= -KNs^2$ from $\hat H_0$ and in first order degenerate perturbation theory $E_s=-KNs^2-\lambda (N-1)s^2+\lambda s^2=(-K-\lambda)Ns^2+2\lambda s^2$ and are split from the first excited level, which requires the introduction of a soliton anti-soliton pair, by an energy of $4\lambda$.  In each order of perturbation theory less than $2s$, the degenerate multiplet of states mixes with states of higher energy, but due to invariance under translation, the corrections brought to each state are identical and the degeneracy is not split.  However, at order $2s$, the degenerate multiplet is mapped to itself.  This causes it to split in energy and the states to reorganize into a band.  Indeed, $\hat V^{2s}$ contains the term $(S_{k+1}^-S_{k+2}^+)^{2s}$ and    $(S_{k-1}^+S_{k}^-)^{2s}$.  When acting on the ket $\ket k$ flips the anti-aligned pair of spins at positions $k+1,k+2$ and at $k-1,k$ respectively.  It is easy to see that flipping this pair of spins has the effect of translating the soliton $\ket k\rightarrow\ket {k+2}$ and $\ket k\rightarrow\ket {k-2}$ respectively.  All other terms in $\hat V^{2s}$ map to states out of the degenerate subspace, either inserting a soliton anti-soliton pair or changing the value of $S^z$ to non extremal values, and hence do not contribute to breaking the degeneracy.  To compute the splitting and the corresponding eigenstates, we follow \cite{bab1}, we have to diagonalize the  $N\times N$ matrix 
with components $b_{\mu,\nu}$ given by
\bea
b_{\mu,\nu}=\braket{\mu|\hat V\mathcal{A}^{{2s-1}}|\nu}, \quad \mu,\nu = 1, 2,\cdots, N
\eea
where $\mathcal{A}^{{2s-1}}= \lb\frac{\mathcal{Q}}{E_{s}-\hat H_{0}}\hat V\rb^{2s-1}$, and $\mathcal{Q}=1-\sum \ket{\mu}\bra{\mu}$. 
The calculation of the components is straightforward, looking at $b_{\mu,1}$ we find
\begin{align}
b_{\mu,1}&= \lb\frac{\lambda}{2}\rb^{2s}\braket{\mu| S^-_2S^+_3\lb\frac{\mathcal{Q}}{E_{s}-\hat H_{0}}S^-_2S^+_3\rb^{2s-1}|1}\nonumber\\&+\lb\frac{\lambda}{2}\rb^{2s}\braket{\mu| S^+_NS^-_1\lb\frac{\mathcal{Q}}{E_{s}-\hat H_{0}}S^+_NS^-_1\rb^{2s-1}|1}.
\label{eqn24}
\end{align}
Applying the operators $2s$ times on the right hand side we obtain
\begin{align}
b_{\mu,1}&= \mathcal{C}[\braket{\mu|3}+\braket{\mu|N-1}]
\end{align}
where $\mathcal{C}$ is given by
\begin{align}
\mathcal{C}&=\pm\lb\frac{\lambda}{2}\rb^{2s}\prod_{m=1}^{2s} m(2s-m+1)\prod_{m=1}^{2s-1}\frac{1}{K m(2s-m)}\nonumber\\
&=\pm K\lb\frac{\lambda}{2K}\rb^{2s}\bigg[\frac{(2s)!}{(2s-1)!}\bigg]^2=\pm 4Ks^2\lb\frac{\lambda}{2K}\rb^{2s}.\label{c}
\end{align}
The first product in Eqn.\eqref{c} comes from the two square roots that accompany the action of the raising and lowering operators, and the second product is a consequence of the energy denominators. The plus or minus sign arises because we have $2s-1$ products of negative energy denominators in Eq.\eqref{eqn24}, so if $s$ is integer, $2s-1$ is odd and we get a minus sign while for half-odd integer $s$, $2s-1$ is even and we get a plus sign. Similarly, one can show that $b_{\mu,\nu}= \mathcal{C}[\braket{\mu|\nu+2}+\braket{\mu|\nu-2}]$ defined periodically of course. Thus we find that the  matrix, $[b_{\mu,\nu}]$, that we must diagonalize is  a circulant matrix \cite{dav} 
\begin{equation}
[b_{\mu,\nu}] =
 \mathcal{C}\begin{pmatrix}
  0 & 0 & 1&0& \cdots & 1& 0 \\
  0 & 0 &0&1&\cdots & 0&1 \\
  1 & 0 & 0 & 0& 1&\cdots&0 \\
  \vdots  & 1 &0& \ddots&\cdots&\ddots  \\
  1 & \cdots &\ddots & \cdots &0 &0&0 \\
 0 & 1 &\cdots & 1\cdots &0 &0&0
 \end{pmatrix}.
 \label{eqn28}
\end{equation}
In this matrix each row element is moved one step to the right, periodically,  relative to the preceding row. The eigenvalues and eigenvectors are well known.  The $j^{\rm th}$ eigenvalue is given by
\beq
\varepsilon_j = b_{1,1} +b_{1,2}\omega_j +b_{1,3}\omega_j^2+\cdots+b_{1,N}\omega_j^{N-1}
\eeq
where  $\omega_j= e^{i\frac{2\pi  j}{N}}$  is the $j^{\rm th}$ , $N^{\rm th}$  root of unity  with corresponding eignvector $\ket{\frac{2\pi  j}{N}}=(1,\omega_j,\omega_j^2,\cdots ,\omega_j^{N-1})$, for $j=0,1,2,\cdots,N-1$.   For our matrix, Eq.\eqref{eqn28}, the only nonzero coefficients are $b_{1,3}$ and $b_{1,N-1}$, thus the one soliton energy bands are
\begin{align}
\varepsilon_j& = \mathcal{C}(\omega_j^2 +\omega_j^{N-2})= \mathcal{C}(\omega_j^2 +\omega_j^{-2})\nonumber\\&
=2\mathcal{C}\cos\lb\frac{4\pi j}{N}\rb .
\label{eqn31}
\end{align}
Introducing the Brillouin zone momentum $q= j\pi/N$,  the energy bands Eq.\eqref{eqn31} can be written as
 \bea
 \varepsilon_q  
=2\mathcal{C}\cos\lb{4q}\rb
\eea
which is gapless unlike the magnon dispersion in Eq.\eqref{mag} but is doubly degenerate as the cosine passes through two periods in the Brillouin zone.  The exact spectrum is symmetric about the value $N/2$.   With $[x]$  the greatest integer not greater than $x$, the states  for $j=[N/2] -k$ and $j=[N/2] +k+1$ for $k=0, 1,2,\cdots, [N/2] -1$ are degenerate as $\cos\lb\frac{4\pi ([N/2] -k)}{N}\rb= \cos\lb\frac{4\pi ([N/2] +k+1)}{N}\rb$ since $[N/2] =N/2 -1/2$. However the state with $k=[N/2]$ is not paired, only $j=0$ is allowed.  When $s$ is and integer, $\cal C$ is negative and the  unpaired state $j=0$ is the ground state which is then non-degenerate, but for $s$ a half odd integer, $\cal C$ is positive, and the ground states are the degenerate pair with $j=[N/2], [N/2] +1$ in accordance with Kramer's theorem \cite{kram}.  However, in the thermodynamic limit, $N\rightarrow\infty$, the spectrum simply becomes doubly degenerate for all values of the spin and gapless.

{\it  Conclusion-}  We have found the ground state and the low lying spectrum for a periodic spin chain in the limit of large spin,  large $z$-component anisotropy and and weak antiferromagnetic exchange coupling between nearest neighbours.  For even number of sites, we find that the ground state is unique and corresponds to the symmetric or the anti-symmetric superposition of the two fully anti-aligned Néel states.    Then the other combination is split in energy, proportional to $\left(\frac{\lambda}{2 K}\right)^{sN}$.    We find this result through an instanton using the spin coherent state path integral.  Thus in the thermodynamic limit, the two Néel states are the degenerate ground states, actually allowing for long range order.  However, there is no spontaneous symmetry breaking, there is explicit symmetry breaking as the $z$-component anisotropy explicitly breaks the rotational invariance.  There is no massless excitation.  The first excited state of this system corresponds to the creation of a soliton  anti-soliton pair, with a minimum energy cost of $4\lambda$.  The magnons (spin waves) are very highly gapped, due to the large anisotropy, with a minimum energy cost $\sim K$.  For an odd number of sites the situation is markedly different.  There is no fully aligned Néel state as the system is frustrated.  The chain must contain at least one soliton. The soliton can be up-up or down-down giving a total $z$ component of spin $s$ or $-s$ respectively.  Since the $z$-component of the spin is conserved, theses states are in orthogonal super-selection sectors.   As the position of the soliton is arbitrary, the ground state in each sector is nominally $N$ fold degenerate.  Perturbation to the order $2s$ mixes these states into each other, breaking the degeneracy and creating a gapless band and destroying the possibility of long range order.  In the thermodynamic limit, the ground state is doubly degenerate in each sector.

{\it  Acknowledgments-} We thank  NSERC of Canada for financial support. 


 \end{document}